# A Graphene Field Effect Device

M.C. Lemme, *Senior Member, IEEE*, T.J. Echtermeyer, M. Baus, and H. Kurz

## Abstract

In this letter, a top-gated field effect device (FED) manufactured from monolayer graphene is investigated. Except for graphene deposition, a conventional top-down CMOS-compatible process flow is applied. Carrier mobilities in graphene pseudo-MOS structures are compared to those obtained from top-gated Graphene-FEDs. The extracted values exceed the universal mobility of silicon and silicon-on-insulator MOSFETs.

Max C. Lemme and Heinrich Kurz are with the Advanced Microelectronic Center Aachen (AMICA), AMO GmbH, Otto-Blumenthal-Str. 25, 52074 Aachen, Germany (phone: +49 241 8867 207, email: lemme@amo.de)

Tim J. Echtermeyer, Matthias Baus and Heinrich Kurz are with the Institute of Semiconductor Electronics, RWTH Aachen University, Sommerfeldstr. 24, 52074 Aachen, Germany

Financial support by the German Federal Ministry of Education and Research (BMBF) under contract number NKNF 03X5508 ("ALEGRA") is gratefully acknowledged.

# I. INTRODUCTION

Moore's law, the scaling rule of thumb turned dogma, has dictated ambitious innovation cycles in silicon technology over the last four decades [1][2]. Along the way, it has provided the fundamental CMOS technology for today's global information society. While the end of silicon has been predicted a number of times for technological reasons it has not only persevered, but is in fact set to remain the driving technology for at least 15 more years. Even beyond this silicon horizon, a demise of CMOS technology is unlikely. Instead, a range of add-on technologies are envisioned to boost the silicon workhorse.

One of the most promising future options to enhance silicon is the introduction of carbon-based electronics [3]. In recent years, intriguing electrical properties have been found in carbon nanotubes (CNTs) [4]. The major disadvantage of CNTs, however, is their random distribution, which clearly hampers their utilization as a replacement for silicon as a substrate. This leaves two options for carbon-electronics: either self-organization methods for CNTs or carbon "substrates", thin layers with similar properties to CNTs.

Two-dimensional carbon sheets of single and few layers (graphene) have only recently been demonstrated to be thermodynamically stable [5]. Monolayer graphene consists of sp²-bonded carbon atoms arranged in a dense honeycomb crystal structure. It is a semi-metal with an extremely small overlap between the valence and the conduction band (zero-gap material). In its three-dimensional graphite structure graphene sheets are weakly coupled between the layers with van der Waals forces.

The two-dimensional nature of graphene has been confirmed by experimental observation of the quantum Hall effect [6]. Excellent electronic properties with reported carrier mobilities between 3000 and 27000 cm²/Vs make it an extremely promising material for future nanoelectronic devices [5][7]. The carrier transport in graphene takes place in the π-orbitals perpendicular to the surface [8] and the extraordinary transport properties have been attributed to a single spatially quantized subband populated by electrons with a mass of $m_e \approx 0.06\ m_0$ or by light and heavy holes with masses of $m_h \approx 0.03\ m_0$ and $m_h \approx 0.1\ m_0$ [5]. With a mean free path for carriers of L = 400 nm at room temperature, ballistic devices seem feasible, even at relaxed feature sizes compared to State-of-the-Art CMOS technology. The major advantage of graphene over carbon nanotubes is its planar form, which generally allows for highly developed top-down CMOS-compatible process flows.

So far, experimental data has been mainly obtained from mono- or few layer graphene on oxidized silicon wafers (Graphene on Insulator) or decomposed intrinsic silicon carbide [5]-[7]. Here, so-called pseudo-MOS structures have been investigated where the surface of the graphene has been left uncovered. This is not a realistic device situation since a graphene transistor would require an insulator and an electrode on top of the graphene. In contrast to previous work on back-gated graphene, a top-gated graphene field effect device (Graphene-FED) is presented in this letter. The effect of the top-gate on carrier transport is studied. In addition, the carrier mobility in graphene is compared to the universal mobility of silicon and to literature data of ultra-thin body silicon on insulator (SOI) devices.

## II. Experiment

P-type silicon wafers (100) with a boron doping concentration of $N_A = 10^{15}\ cm^{-3}$ have been thermally oxidized to an $SiO_2$ thickness of $t_{ox} = 300$ nm. Graphene has then been deposited onto the silicon dioxide according to the method described in [5] and visually inspected to identify a suitable few layer graphene flake. Titanium (Ti) / gold (Au) contacts have been evaporated after optical lithography and structured by lift-off. Next, electron beam lithography has been used to define a gate electrode on top of the graphene. Finally, a gate stack of silicon dioxide (20 nm), Ti (10 nm) and Au (100 nm) has been evaporated followed once again by a lift-off process. A scanning electron microscope image of the field effect device is shown in

Fig. 1. The graphene flake has a total length from source to drain of L = 7.3 µm, a width of W = 265 nm at the gate region and a gate length of L = 500 nm. An FLG thickness of t = 1.5 nm has been determined by atomic force microscopy after electrical characterization. Raman spectroscopy of the FED (not shown) has identified the presence of a single graphene layer, as it exhibits the characteristic trait of monolayer graphene first published in [9].

## III. Results and Discussion

Fig. 2 compares the back-gate transfer characteristics (log $I_d$-$V_{bg}$) of the Graphene-FED before and after fabrication of the top-gate. A constant source-drain voltage of $V_{ds}$ = 100 mV has been applied and the back-gate field has been swept from $E_{bg}$ = -3.5 MV/cm to $E_{bg}$ = 3.5 MV/cm while measuring the drain current through the graphene layers. Without a top-gate, the electrical field applied by the back-gate modulates the drain current by almost one order of magnitude (Fig. 3, black dots). Potentially, the gate modulation can be improved by using nanoribbons [10] or bilayer graphene [11] with a larger band gap than the semi-metallic monolayer graphene. While ambipolar behavior is observed, hole conduction is favored over electron conduction: negative back-gate fields result in higher drain current modulation compared to positive back-gate fields. The reason is not clearly identified, but may be attributed to chemical doping by adsorbants during processing and handling of the sample [12]. This is also believed to cause the shift of the current minimum towards positive $E_{bg}$.

Carrier mobilities have been calculated based on a Graphene-FED width of W = 265 nm (Fig. 1) using the equation $n_s = \varepsilon_{ox} * V_G / (q * t_{ox})$ for determining the charge carrier density $n_s$, with silicon dioxide permittivity $\varepsilon_{ox}$, gate voltage $V_G$, electron charge q and silicon dioxide thickness $t_{ox}$. A value of $\varepsilon_g$ = 2.4 has been used for the dielectric constant of graphene [13]. Furthermore, the unsymmetrical position of the minimum current in Fig. 2 has been taken into account. The charge carrier mobilities in the uncovered graphene at an effective field of $E_{eff} = \varepsilon_{SiO2}/\varepsilon_g * E_{bg} = 0.4$ MV/cm are estimated to be $\mu_h \geq 4790$ cm$^2$/Vs and $\mu_e \geq 4780$ cm$^2$/Vs.

The lower curve in Fig. 2 shows the back-gate modulation after the top-gate deposition. The top-gate potential has been kept floating during this measurement. A considerable decrease in drain current is observed compared to uncovered graphene. In addition, the current modulation through the back-gate is reduced by the top-gate.

After the deposition of the top-gate, mobility values of $\mu_h = 710$ cm$^2$/Vs and $\mu_e = 530$ cm$^2$/Vs at $E_{eff} = 0.4$ MV/cm and room temperature have been achieved. This reduction compared to uncovered graphene is attributed to the participation of the top π-orbitals to van-der-Waals bonds to the silicon dioxide. This is in accordance with [8], where it was proposed that graphene π-orbitals contributing to van der Waals' bonds have less overlap and thus result in reduced conductivity. Compared to silicon transistors with their universal mobilities of $\mu_h = 95$ cm²/Vs and $\mu_e = 490$ cm²/Vs at 0.4 MV/cm [14] these values are very promising for CMOS applications. Please note that extracted graphene FED mobilities exceed silicon universal mobility over the entire measured range of $E_{eff} = 0$ to 1 MV/cm, particularly in the case of holes. In comparison with ultra thin body silicon on insulator transistors, the values are even more encouraging. Here, hole mobility has been found to drop below $\mu_h \sim 60$ cm$^2$/Vs in $t_{Si} = 3.7$ nm films [15][16] and electron mobility has been found to drop below $\mu_e = 70$ cm$^2$/Vs in $t_{Si} = 2.5$ nm films [17], both in (100) silicon at room temperature.

The top-gate transfer characteristics of the Graphene-FED are shown in Fig. 3 for three different back-gate fields $E_{bg}$. The drain current $I_d$ is clearly modulated by the top-gate field $E_{tg}$. For increasing negative top-gate fields, a constant increase in hole current is observed. For positive top-gate fields, however, there is a distinct plateau between ~0.1 and ~0.4 MV/cm, presumably due to oxide defects. The back-gate field $E_{bg}$ induces an offset of the top-gate transfer characteristics without changing the ambipolar signature. This is attributed rather to a series resistance modulation in the graphene leads next to the top-gate rather than a modulation of the FED channel resistivity alone.

## IV. Conclusion

In this letter, a top-gated transistor-like field effect devices manufactured from monolayer graphene is investigated - to the best of our knowledge - for the first time. Compared to pseudo-MOS structures with uncovered graphene, an additional "standard" transistor gate with $SiO_2$ dielectric reduces electron and hole mobility. Despite the limiting effect of the top-gate electrode, carrier mobilities have been extracted that clearly exceed universal mobility in silicon and even more literature data of ultra-thin body SOI MOSFETs. Furthermore, the top-gate has been used to modulate the drain current, confirming that the field effect resulting from a combined action of top- and back-gate can be applied to graphene devices. While band gap tuning [10][11] will be mandatory to improve device characteristics, this letter confirms the impressive potential of graphene for future electronic devices.

## Acknowledgment

M.C. Lemme thanks L. Risch (Qimonda, Dresden) for fruitful discussions and his encouragement for this exciting topic. The authors thank J. Bolten and T. Wahlbrink (AMO, Aachen) for their e-beam lithography support and M. Ramsteiner (Paul-Drude-Institut, Berlin) for Raman spectroscopy.

# FIGURE CAPTIONS

Fig. 1: Scanning electron microscope image of a graphene transistor.

Fig. 2: Back-gate transfer characteristics of the Graphene-FED with and without top-gate.

Fig. 3: Top-gate transfer characteristics of the Graphene-FED for different back-gate fields $E_{bg}$.

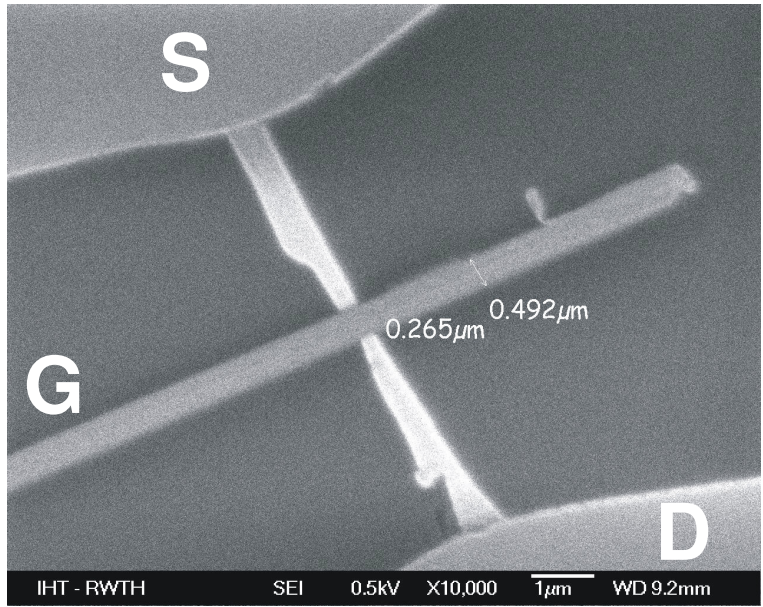

Fig.1

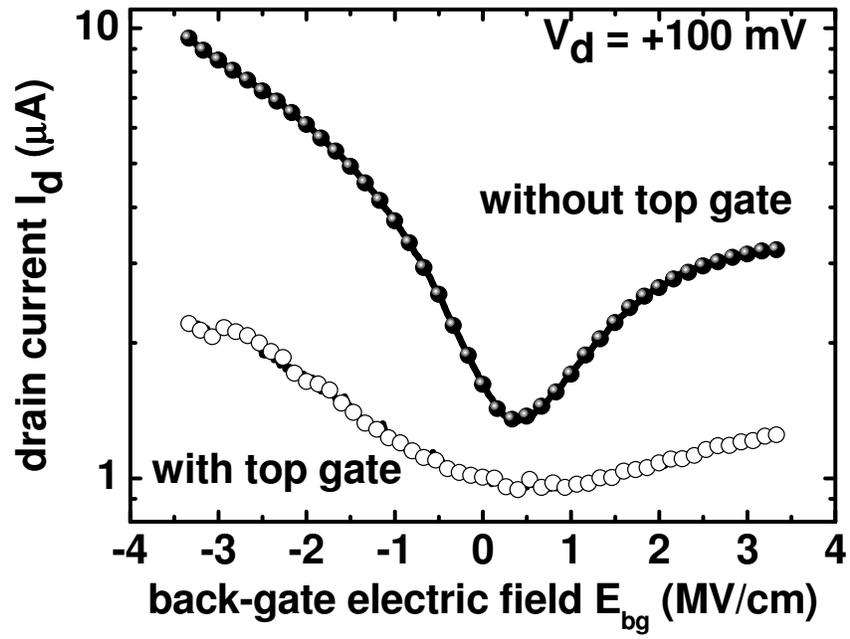

Fig.2

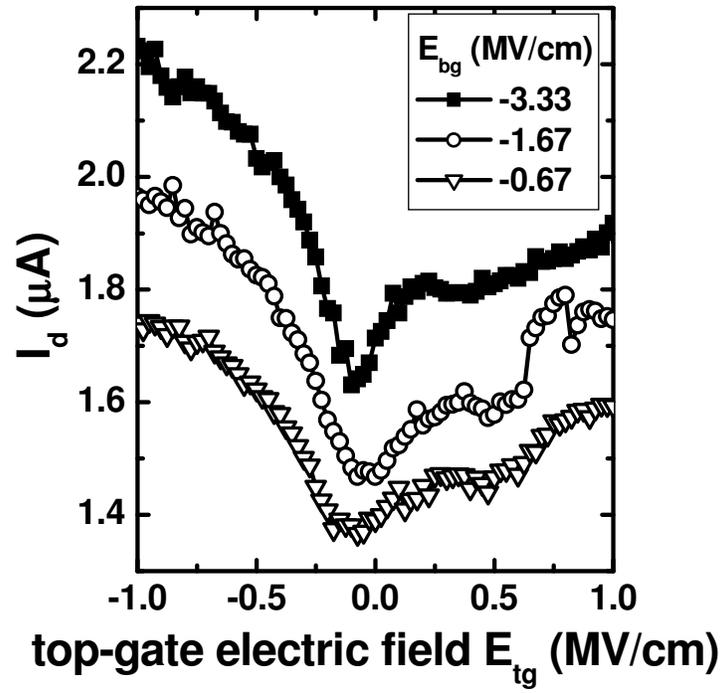

Fig. 3